\documentclass[twocolumn,showpacs,prl]{revtex4}

\usepackage{bm}

\usepackage{graphicx}

\usepackage{subfigure}

\begin{document}

\title{Spectroscopy of Strong-Pulse Superradiance in a Bose-Einstein condensate}

\author{Nir Bar-Gill, Eitan E. Rowen and Nir Davidson}

\affiliation{Weizmann Institute of Science, 76100
Rehovot, Israel}

\date{\today }


\begin{abstract}
We study experimentally superradiance in a Bose-Einstein condensate using a two-frequency pump beam. By controlling the frequency difference
between the beam components, we measure the spectrum of the backward (energy-mismatched) superradiant atomic modes. In addition, we show that
the populations of these modes display coherent time-dynamics. These results are compared to a semi-classical model based on coupled
Schroedinger-Maxwell equations.
\end{abstract}

\maketitle

The effect of superradiant light emission from a collectively excited sample of atoms, first described by Dicke \cite{PhysRev.93.99}, has recently raised renewed interest in the context of Bose-Einstein condensates (BEC), where the coherence properties of both light and matter waves, the spectroscopic properties of the superradiant modes, and their coherent dynamics can be directly explored \cite{S.Inouye07231999,DominikSchneble04182003,PhysRevA.69.041601,PhysRevA.69.041603,PhysRevLett.94.083602,PhysRevA.71.033612}. In the basic BEC superradiance experiments, an elongated BEC was pumped by a beam perpendicular to its long axis. Due to the large optical density along the long axis, the spontaneous emission of BEC atoms excited by the pump beam is amplified along this axis. This creates the superradiant radiation modes, referred to as the endfire modes. The ultracold atoms which undergo the process of absorption and emission recoil at $\pm 45^\circ$ angles compared to the pump beam direction. We refer to these atomic modes as the forward modes.

Following the first observation of superradiance in a BEC \cite{S.Inouye07231999}, off-resonant superradiant modes (backward modes) were measured in experiments with short-pulse excitations \cite{DominikSchneble04182003}. The backward modes result from processes in which BEC atoms re-absorb a photon from the superradiant emission, and emit a stimulated photon into the pump laser. The superradiant light is created through forward scattering events, and therefore these photons have a frequency smaller by twice the atomic recoil energy compared to the original pump frequency. Thus the backward scattering processes are off-resonant by four times the recoil energy. In \cite{DominikSchneble04182003}, the short duration of the pulse allowed for the buildup of these energy-mismatched modes, preventing study of the time dynamics and the spectral content of this process.

These experiments were followed by several theoretical works \cite{PhysRevLett.91.150407,Zobay2006} addressing the coherence buildup, populations and spatial dependence of the superradiant modes. In particular, it was suggested that in the regime where both forward and backward peaks are created, these modes are number squeezed, due to the quantum nature of the process by which they are created \cite{PhysRevLett.91.150407}. In this excitation scheme a zero momentum two-photon excitation creates pairs
of modes with correlated atomic momentum \cite{durr:052707}.


In this letter we study experimentally superradiant Rayleigh scattering in a BEC using a pump laser composed of two frequencies. This allows us to excite both forward and backward superradiant modes using long pulses by setting the difference between these two frequencies to be close to the two-photon resonance. Hence the buildup of the backward peaks compared to the standard forward peaks can be studied spectroscopically and as a function of time. We measure coherent oscillations in the time-dynamics of the superradiant modes. For weak-pulse superradiance we find a single peak two-photon spectrum, centered around $4\omega_R / 2 \pi$, where $\omega_R = \frac{\hbar k_l^2}{2 M}$  is the recoil frequency of the
atom, as expected from energy conservation considerations. For strong-pulse superradiance we find a multi-peaked structure of the two-photon spectrum, and a downward frequency shift of the peak response. Our technique of frequency dependent atomic mode population measurements provides the first kHz-scale spectroscopic data of superradiant processes. A semi-classical model based on coupled Schroedinger-Maxwell equations
\cite{Zobay2006} is used to verify and supplement our experimental data.

\begin{figure}[tbh]
\centering
\subfigure[]{
\includegraphics[width=0.78\linewidth,height=0.5\linewidth]{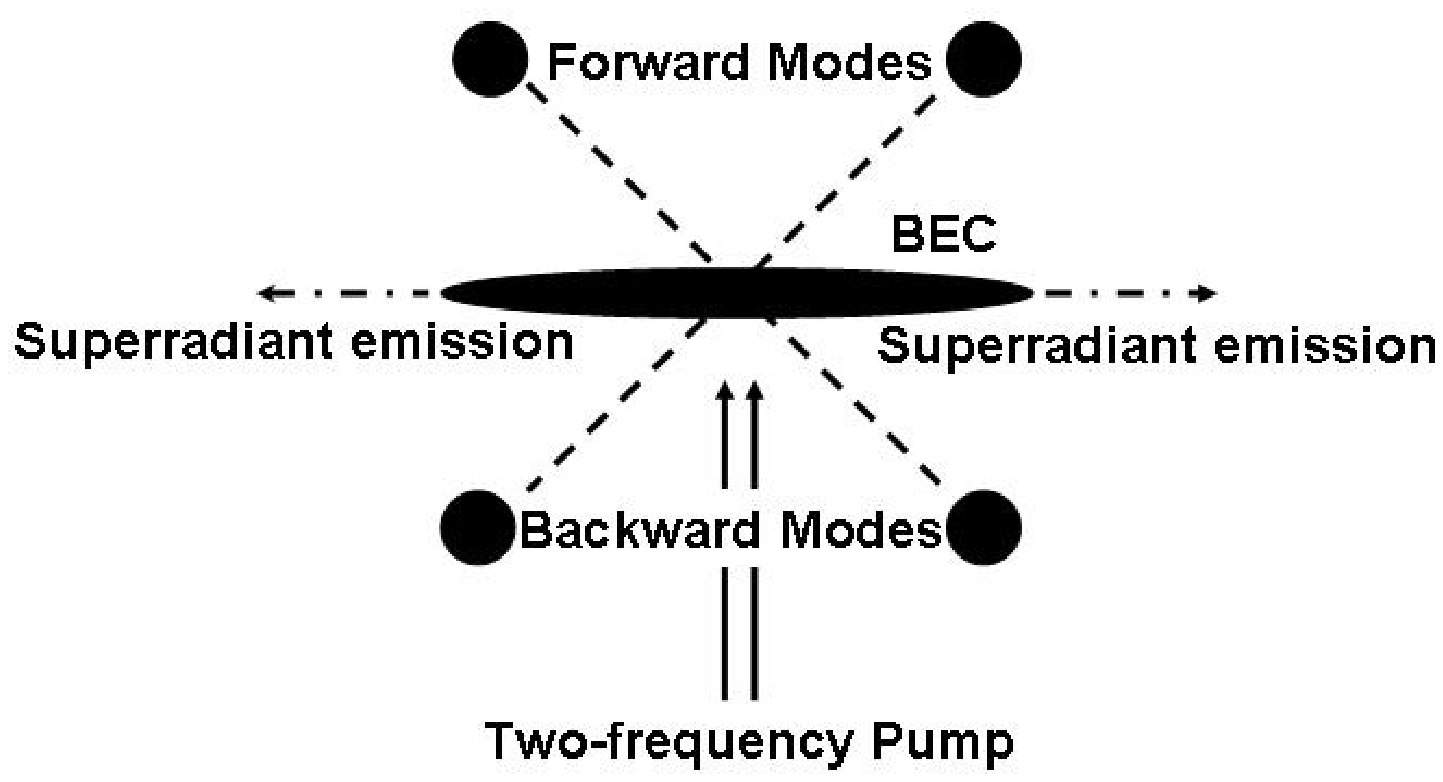}}
\subfigure[]{
\includegraphics[width=0.48\linewidth,height=0.3\linewidth]{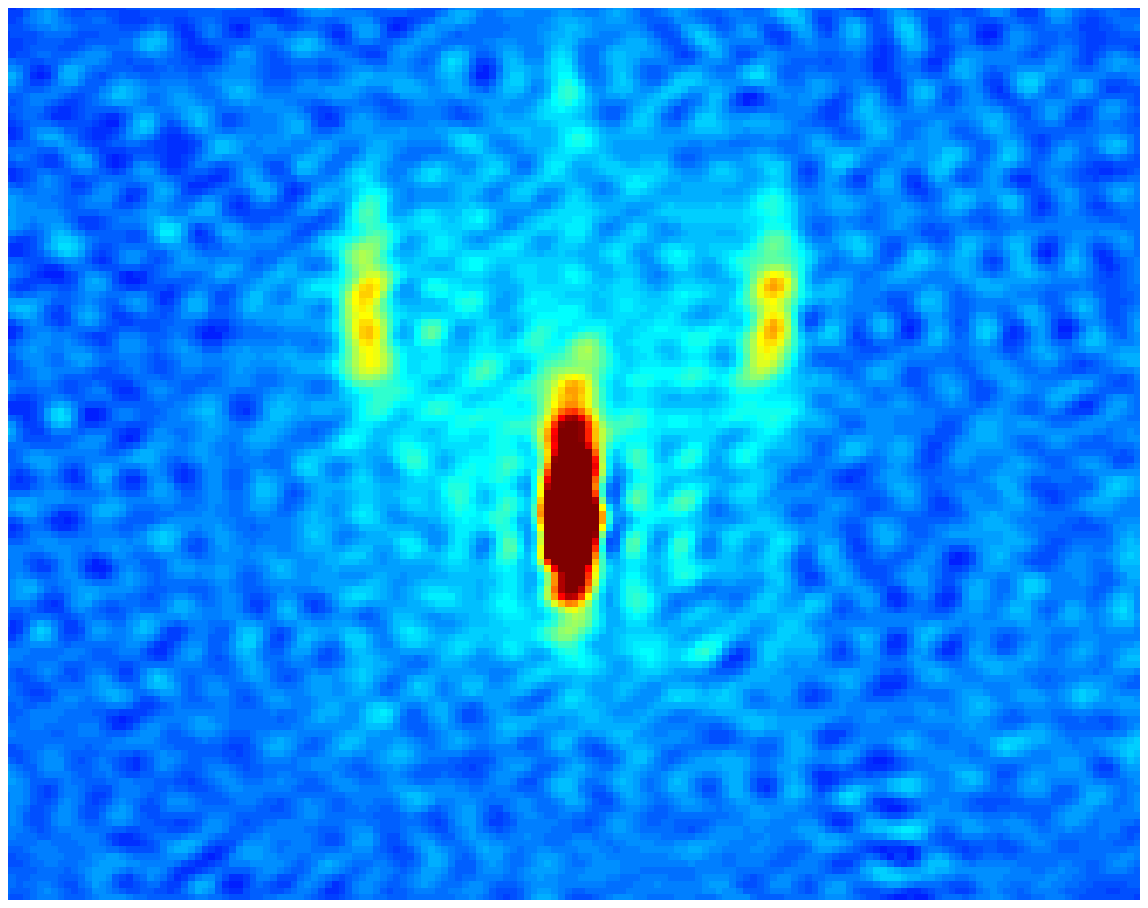}}
\subfigure[]{
\includegraphics[width=0.48\linewidth,height=0.3\linewidth]{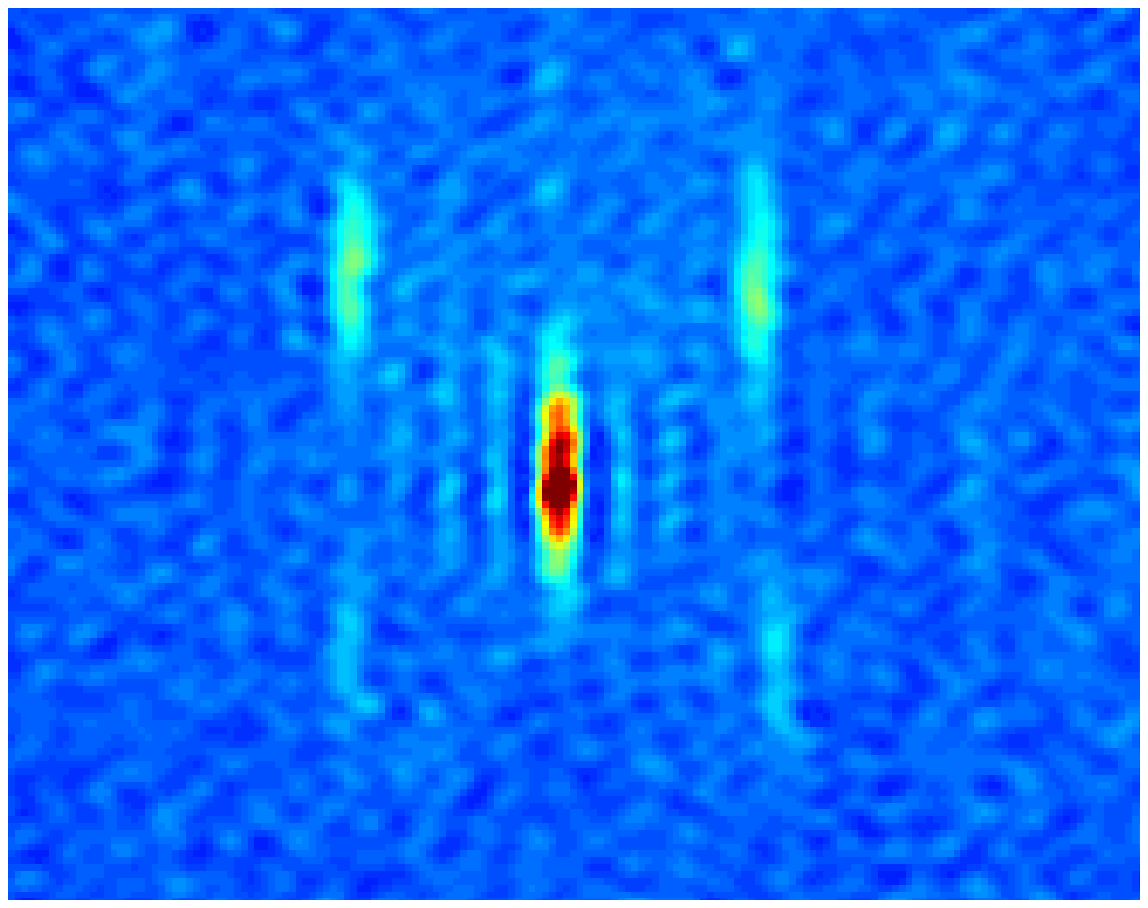}}
\caption{(a) Experimental setup: superradiant emission builds up along the BEC's long axis (endfire modes - see \cite{S.Inouye07231999}), creating forward modes of scattered atoms. The backward atomic modes result from stimulated absorption of superradiant photons and emission into the pump laser beam. (b) Absorption image after $38$ msec time-of-flight of forward superradiance modes obtained with a single-frequency pump and (c) an X
pattern (forward and backward superradiance modes) obtained for a two-frequency pump with $ f=4\omega_R / 2 \pi=15$ kHz, both for a pulse
duration of $200 \ \mu$sec. Figure size is 1.1 by 0.8 mm.} \label{fpics}
\end{figure}

The experimental setup (Fig. \ref{fpics}(a)) consists of a nearly pure $^{87}$Rb BEC of $2 \cdot 10^5$ atoms in the $|F=2,m_f=2 \rangle$ ground state
trapped by a quadrupole-Ioffe magnetic trap with radial and axial trapping frequencies of $2 \pi \times 350$ Hz and $2 \pi \times 30$ Hz, respectively, and a chemical potential
of $\mu / h \simeq 5$ kHz. These parameters result in an elongated BEC, with a length of $l \simeq 70 \mu$m, and a diameter of $d \simeq 7 \mu$m. Such an elongated geometry is required in order to create well-defined superradiant emission modes, for which the gain of the superradiance process is largest. The excess gain of these modes is measured by the Fresnel number $F=\pi d^2/4 l \lambda \simeq 0.7$, indicating that the large gain selects very few superradiant modes. The BEC is excited from below by a laser beam of intensity I $\simeq 30 \ \mu$W/cm$^2$, red-detuned by 130 MHz from the $|F=2,m_f=2 \rangle$ to
$|F'=3,m_f=3 \rangle$ transition, whose linear polarization allows for superradiant Rayleigh scattering \cite{S.Inouye07231999}. These intensity and detuning result in a strong coupling between the radiation and atomic modes \cite{DominikSchneble04182003}. This beam is passed through an acousto-optic modulator, which is modulated by function generators at two frequencies to create a two-frequency pump. The two frequencies are created by driving a large carrier frequency (30 MHz), and then modulating the amplitude at a controlled frequency. The modulation is strong enough to deplete the carrier, resulting in two, equally strong side-bands at the desired frequency difference. This accurately controlled frequency difference allows for spectroscopic measurements of the backward and forward superradiant modes. Measurements are carried out by imaging the atoms 38 msec after turning of the magnetic trap (Time Of Flight absorption imaging, see Fig. \ref{fpics}). Thus we do not measure the superradiant light itself, but rather the excited superradiant atomic modes.

We use a pump pulse duration of $200 \ \mu$sec for which superradiant dynamics occurs during the pumping process, including induced emission and absorption, which account for the off-resonant backward atomic modes. Therefore, following the nomenclature introduced in the original superradiance papers \cite{PhysRevA.11.1507}, our experiment includes also superfluorescence and amplified spontaneous emission effects. Our theoretical model includes all these effects in a semi-classical framework \cite{Zobay2006}.


Fig. \ref{fpics}(a) depicts schematically the excitation process and the superradiant emissions, as well as the relative momentum-space
distribution of the atomic superradiant modes. 
We also show typical absorption images of the excited BEC after a 38 msec time of flight, for an excitation with a single frequency pump (b) and a two-frequency pump with a frequency difference $f=4\omega_R / 2 \pi=15$ kHz (c).
As can be seen, the $200 \ \mu$s pulse is long enough to exclude population of the backward superradiant modes for the single-frequency pump, but the two-frequency pump yields clearly visible backward modes. Since the superradiance modes are spatially separated in the time of flight images, the population of each mode can be readily quantified.

In order to theoretically model the experiment, we generalize the
formalism of \cite{Zobay2006} to include a two-frequency pump beam. This is a 1D model, which describes the pump and superradiance modes $e_\pm$ using Maxwell's equations, and their coupling to the atomic modes $\psi_{nm}$, governed by Schrodinger's equation. In this formalism the slowly-varying amplitude approximation is used. Since
the difference between the two pump frequency components is on the order of $\omega_R /2 \pi << k_l c$, we can simply introduce an additional pump term with a time dependence of $e^{i 2 \pi f t}$, where $f$ is the frequency difference. Thus, the coupled evolution equations of the atomic modes (normalized to 1) are given, using the dimensionless coordinates $\tau = \omega_R t$ and $\xi = k_l z$, by
\begin{eqnarray}
\lefteqn{i \frac{\partial \psi_{nm} (\xi,\tau)}{\partial \tau} = - \frac{1}{2} \frac{\partial^2 \psi_{nm}}{\partial \xi^2} - i m \frac{\partial \psi_{nm}}{\partial \xi}} \nonumber \\
&+& \sqrt{\frac{G}{N}} \sqrt{\frac{c k_l}{2 \omega_R}} \left[ \left(a_1 + a_2 e^{i 2 \pi f \tau} \right) e_{+}^{*} \psi_{n-1,m+1} e^{i (n-m-2) \tau} \right. \nonumber \\
&+& \left. \left(a_1 + a_2 e^{i 2 \pi f \tau} \right) e_{-}^{*} \psi_{n-1,m-1} e^{i (n+m-2) \tau} \right. \nonumber \\
&+& \left. \left(a_1^{*} + a_2^{*} e^{-i 2 \pi f \tau} \right) e_{+} \psi_{n+1,m-1} e^{-i (n-m) \tau} \right. \nonumber \\
&+& \left. \left(a_1^{*} + a_2^{*} e^{-i 2 \pi f \tau} \right) e_{-} \psi_{n+1,m+1} e^{-i (n+m) \tau} \right], \label{atomeq}
\end{eqnarray}
and the equation for the radiation modes are
\begin{eqnarray}
\lefteqn{e_{+} (\xi,\tau) = - i \sqrt{GN} \sqrt{\frac{2 \omega_R}{c k_l}} \int_{-\infty}^{\xi} d \xi'} \\
&& \sum_{(n,m)} e^{i (n-m) \tau} \psi_{n,m} (\xi',\tau) \psi_{n+1,m-1}^{*} (\xi',\tau) e^{-\tau/\tau_0} \nonumber \\ \label{radeq} 
\lefteqn{e_{-} (\xi,\tau) = - i \sqrt{GN} \sqrt{\frac{2 \omega_R}{c k_l}} \int_{\xi}^{\infty} d \xi'} \\
&& \sum_{(n,m)} e^{i (n+m) \tau} \psi_{n,m}
(\xi',\tau) \psi_{n+1,m+1}^{*} (\xi',\tau) e^{-\tau/\tau_0}. \nonumber
\end{eqnarray}
In this notation the indices $n,m$ label the atomic modes. The condensate is labeled $n=m=0$, the forward modes have $n=1,m=\pm 1$, while the backward modes have $n=-1,m=\pm 1$. $a_1$ and $a_2$ are the amplitudes of the different pump components (taken to be equal in the rest of this paper). The coupling coefficient is given by
\begin{equation}
G = \frac{3 \pi R N}{4 k_l^2 A \omega_R}, \label{Geq}
\end{equation}
where $R$ is the Rayleigh scattering rate, $N$ is the atom number, and $A$ is the cross-sectional area of the BEC. The growth rate of the superradiance process is governed by $G$.
This rate yields the strong-coupling regime at the limit $G > 1$, for which exponential gain is significant. For $G << 1$ exponential gain is small, which results in the weak-coupling regime. 
We also introduce a phenomenological dephasing term between the atomic modes, to account for the fast radial dynamics of the atoms in the trap (with a time constant of $\tau_0 = 200 \ \mu$sec, due to a radial trapping frequency of $\simeq 2 \pi \times 350$ Hz). Finally, since the model presented in Eqs. (\ref{atomeq}-\ref{radeq}) is semi-classical, it requires an
initial seed of approximately one atom per front mode to replace the initial quantum buildup of the superradiant modes
\cite{Zobay2006}. 

\begin{figure}[tbh]
\centering
\subfigure[]{
\includegraphics[width=0.78\linewidth,height=0.4\linewidth]{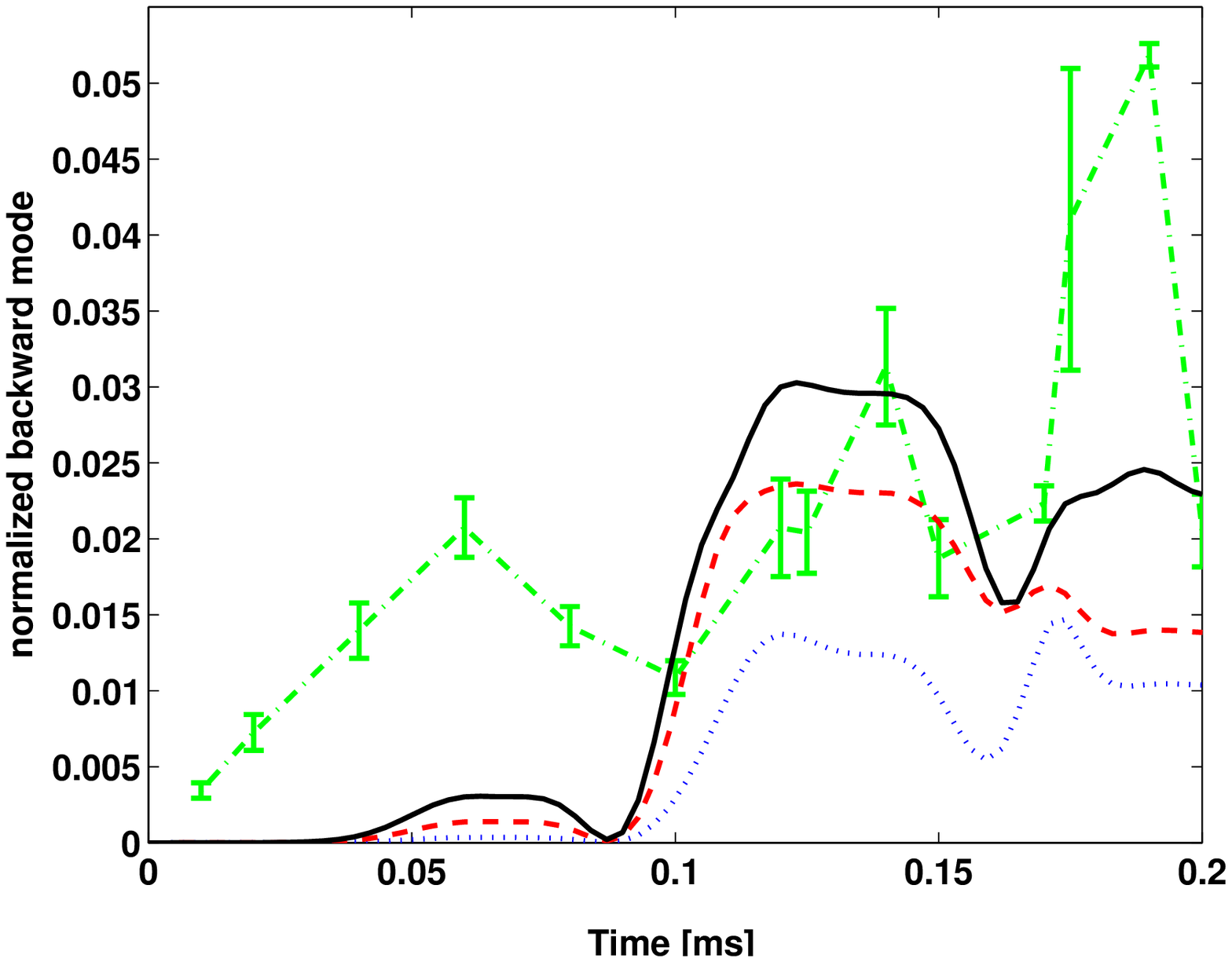}}
\subfigure[]{
\includegraphics[width=0.78\linewidth,height=0.4\linewidth]{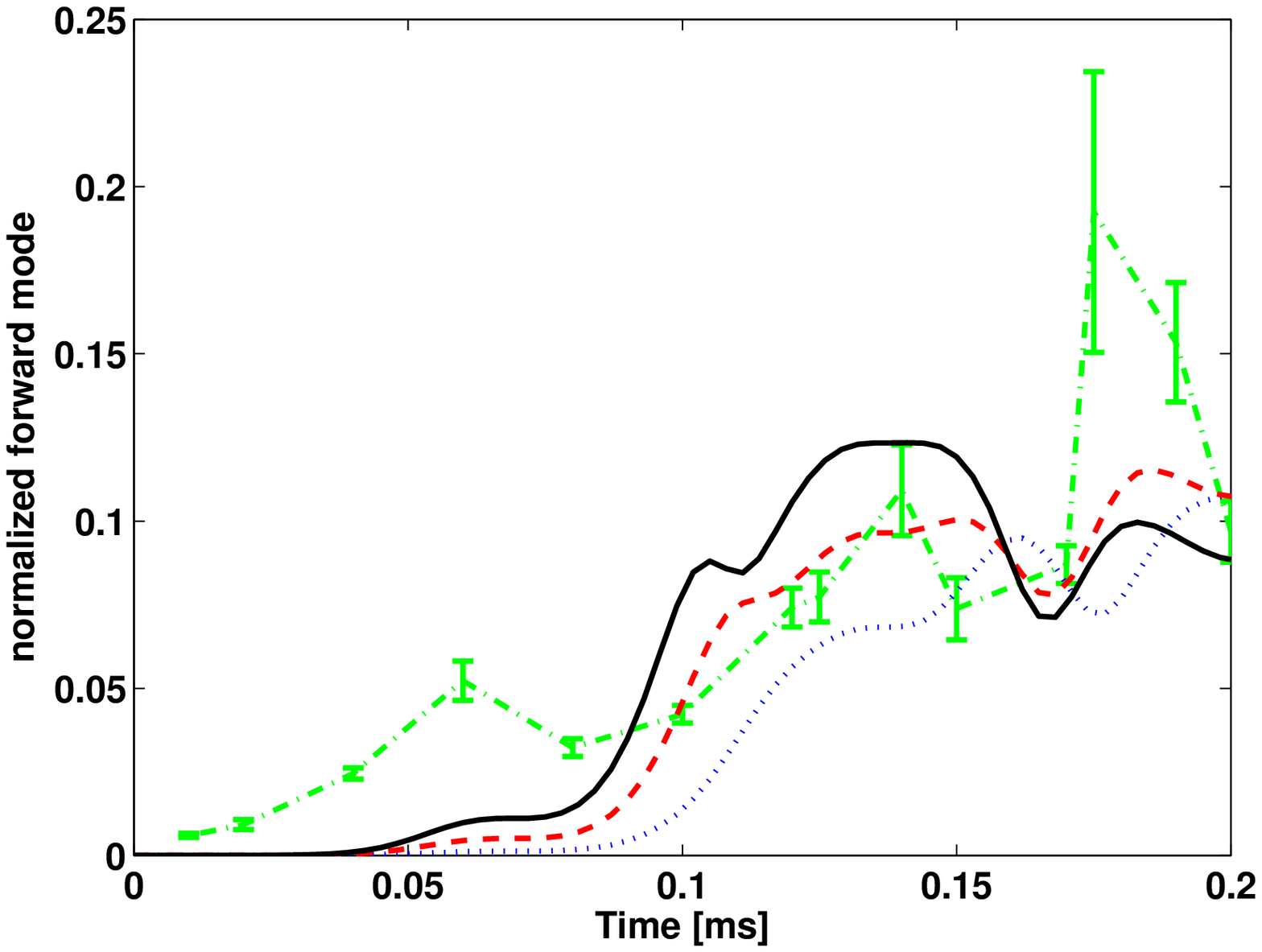}}
\caption{Measurement (green, dashed-dotted) and calculation (seed=0.5 atom: blue, dotted; seed=1 atom: red, dashed; seed=1.5 atoms: black, solid; a factor of 3 in the vertical axis was used) of the backward mode (a) and forward mode (b) populations as a function of time for a pump beam frequency difference of $f = 4
\omega_R / 2 \pi \simeq 15$ kHz. These results are for strong coupling, for a fixed phase of $\pi$ between the frequency components. Error bars indicate the error of the mean.} \label{td_calc_exp}
\end{figure}
In Fig. \ref{td_calc_exp} we plot the measured time dynamics of the backward and forward superradiant mode population, driven by a two-frequency pump with a frequency difference of $ f=4\omega_R / 2 \pi=15$ kHz, at a fixed phase of $\pi$ between the two frequencies. In this and in the following figures we plot the atomic mode population normalized by the total number of atoms - the fraction of atoms in the relevant mode. Oscillations with duration of $60 - 80 \ \mu$sec are evident in the population of both modes, indicating {\em coherent dynamics}. This results from the induced absorption and emission of the superradiant
light during the long duration of the pump pulse. 

Our experimental parameters fit a dimensionless coupling coefficient of $G \simeq 13$ (corresponding to eq. (\ref{Geq}) with the measured values of $N$, $R$, etc.). The calculated time-dynamics of the mode populations for these parameters is also shown in Fig. \ref{td_calc_exp} using an initial seed of 0.5, 1 and 1.5 atoms, respectively, for the population of the forward superradiant modes (no initial seed is used for the backward
modes). 
Despite the dependence of the dynamics on the chosen seed, qualitative agreement exists between the calculated and the experimental results, in particular in the oscillation time-scale. However, the vertical scale for the calculations was scaled by a factor of 3. This discrepancy may be partly due to collisions between the superradiant modes and the BEC, which reduce the measured population of the superradiant modes \cite{katz_beliaev}, but are not included in our model. We note that the theory does not take into account collisions between the superradiant modes and the BEC, which could affect the calculated dynamics and their agreement with the experiment. Our calculations show that the oscillation time-scale is dependent on the pump beam frequency difference rather than on the coupling strength. This indicates that the observed oscillations result from the interference of the atomic modes created by the two pump components. Forward modes are created by each frequency component of the pump beam, and these modes acquire a relative phase proportional to the pump beam frequency difference. This effect creates oscillations in both the forward and backward mode time dynamics, with a frequency given by the frequency difference in the pump beam.

We note that the superradiance process has a strong dependence on the initial seed due to its exponential gain, as can be seen from the calculations in Fig. \ref{td_calc_exp}. This dependence is also evident in the spread of experimental results, measured several times for the same set of parameters. The measured spread can be attributed to the spread in the effective seed in each experimental realization, resulting from either quantum fluctuations \cite{PhysRevLett.39.547,PhysRevLett.42.1740,PhysRevA.20.2047}, or from thermal fluctuations. In our experiment the thermal fraction of the atoms is unobservable. However, our imaging system is not sensitive to small atom numbers (less than $\sim 100$), and therefore at present we cannot rule out the existence of thermal seeding. 

\begin{figure}[tbh]
\centering
\subfigure[]{
\includegraphics[width=0.48\linewidth,height=0.3\linewidth]{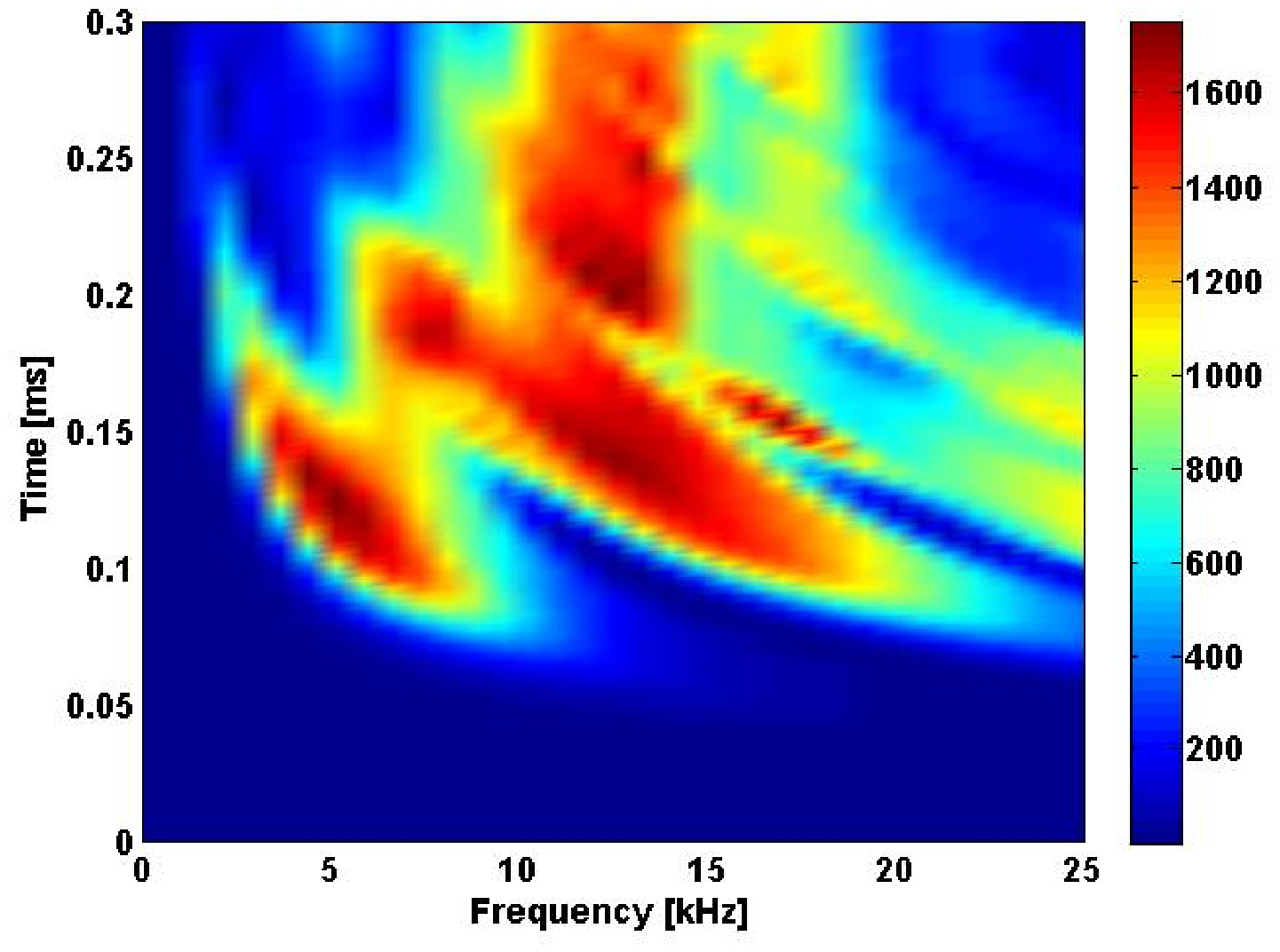}}
\subfigure[]{
\includegraphics[width=0.48\linewidth,height=0.3\linewidth]{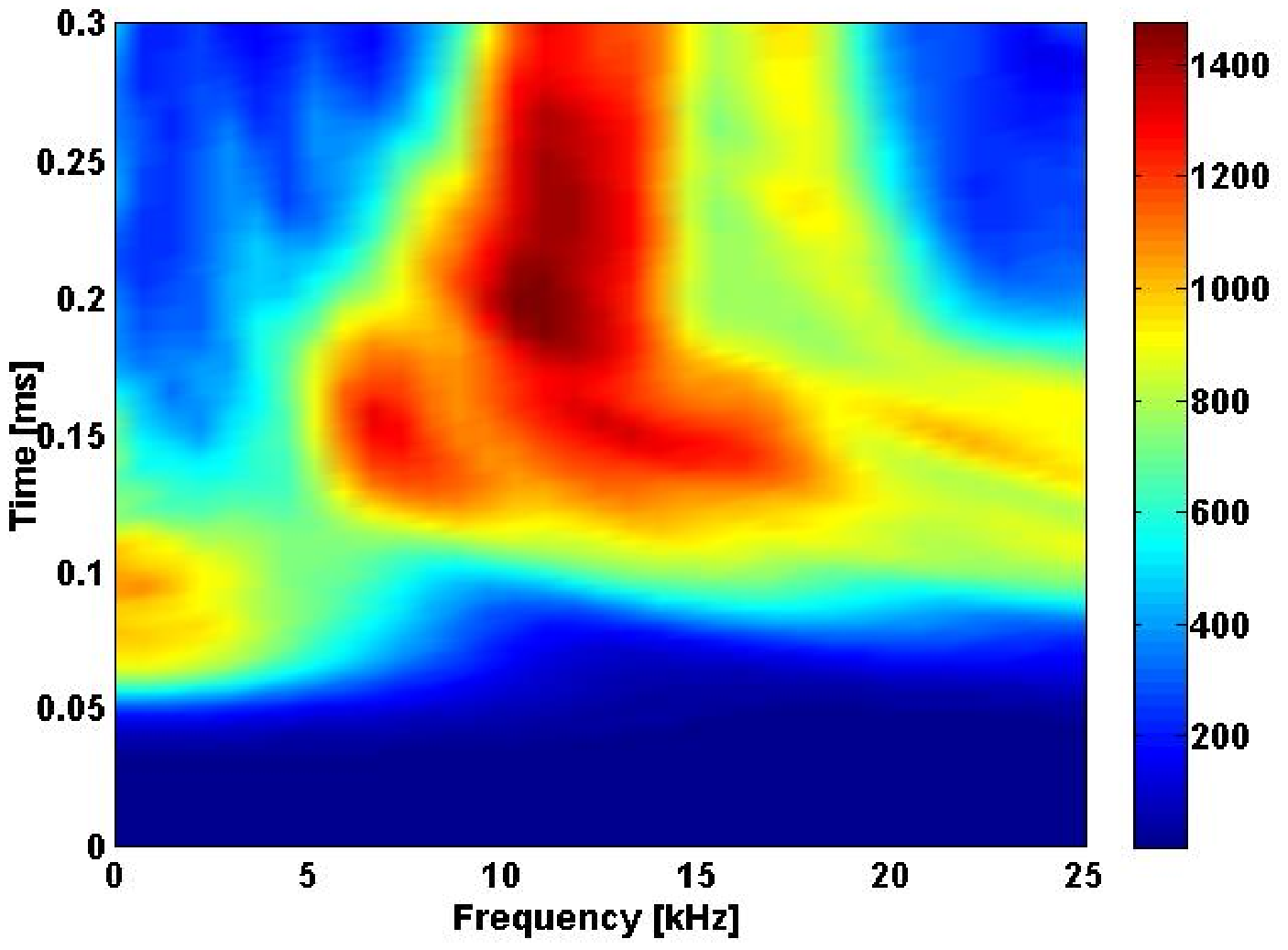}}
\subfigure[]{
\includegraphics[width=0.78\linewidth,height=0.4\linewidth]{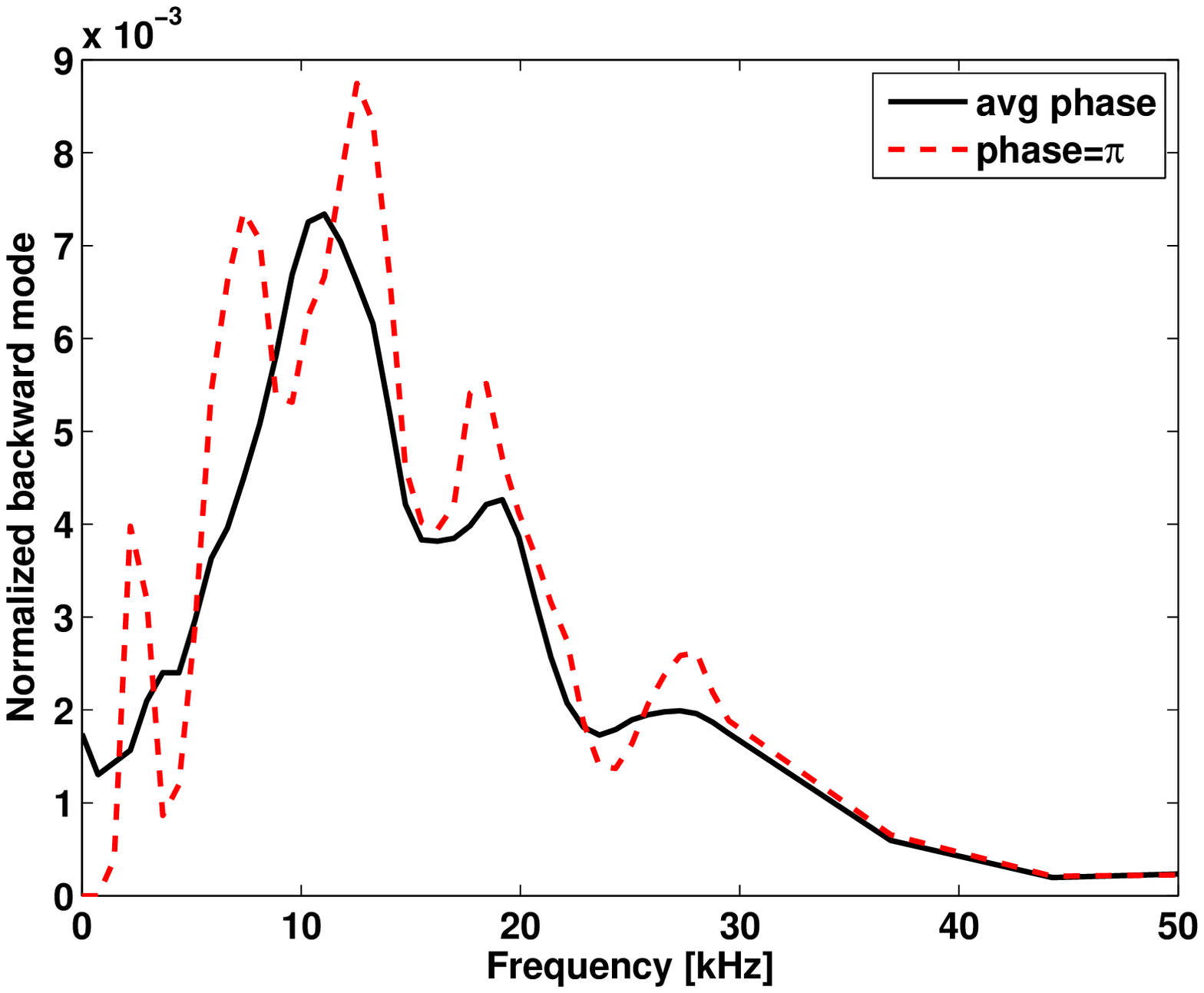}}
\subfigure[]{
\includegraphics[width=0.78\linewidth,height=0.4\linewidth]{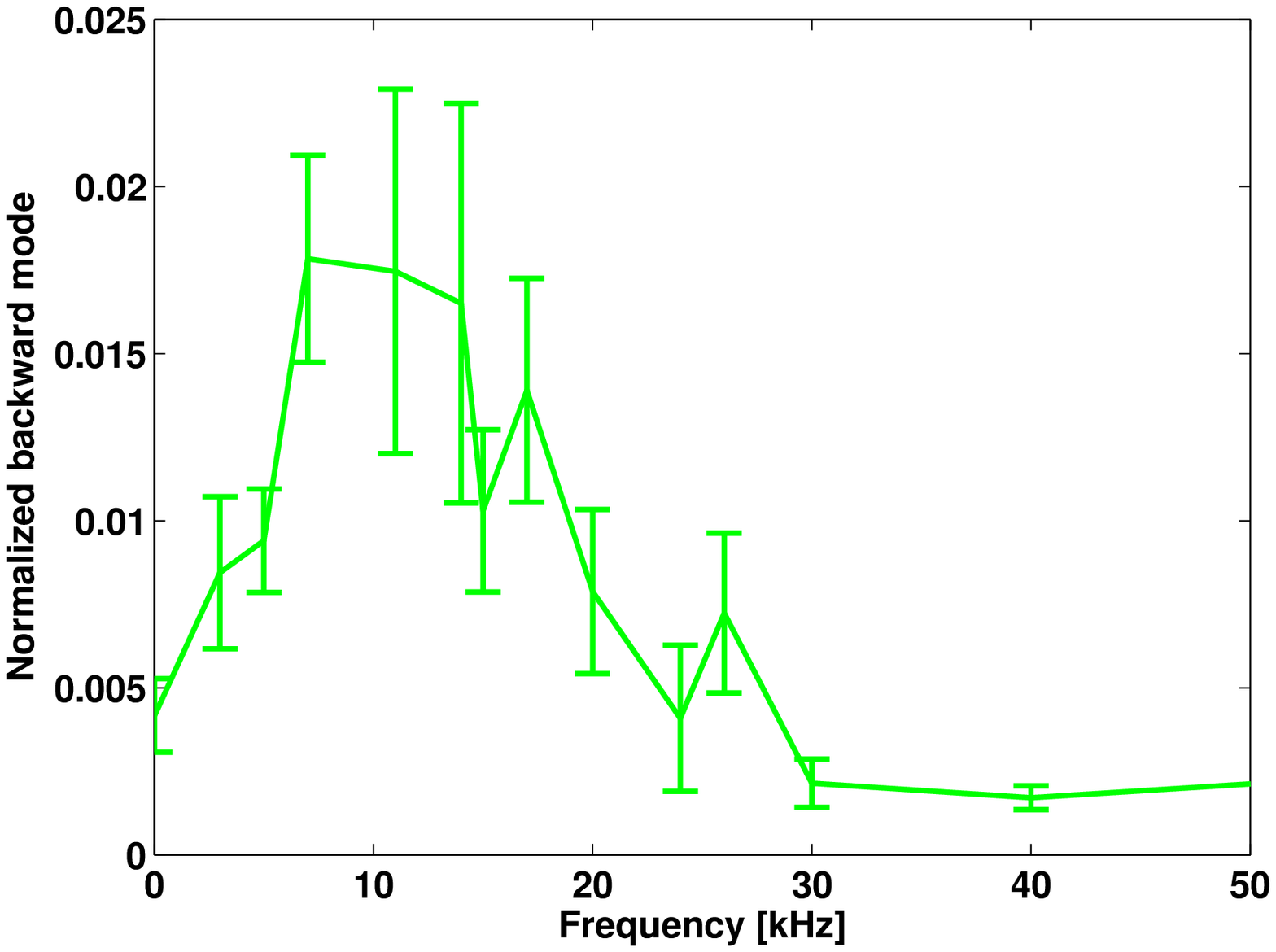}}
\caption{Comparison between experimental data and theoretical calculations for the spectroscopic response of the backward modes in the strong
coupling regime. (a) Calculation of the backward modes population as a function of time and of the pump beam frequency difference for a fixed
phase of $\pi$, and (b) averaged over 10 randomly chosen phases between the two pump frequencies (using a seed of 1 atom). (c) Calculated backward mode response for an averaged phase and for a fixed phase of $\pi$, shown at $t = 200 \ \mu$sec, as a function of the
frequency difference. (d) Experimental measurement of the spectroscopic response of the backward modes for a fixed phase pump. Both experiment
and theory show a broad response centered at $\sim 13$ kHz with a width of $\sim 20$ kHz. The fixed phase calculation shows a structure
consisting of several peaks.} \label{spect_strong_exp_theory}
\end{figure}
Next, we continue our study to include the dependence of the dynamics on the spectral content of the pump. In Fig. \ref{spect_strong_exp_theory} the calculated response of the backward mode in the strong pulse regime is shown as a function of both time and frequency difference, (a) for a fixed phase of $\pi$ between the two pump frequencies  and (b) averaged over random phases. Coherent dynamical oscillations are evident for the fixed-phase calculation, while the average phase result shows a smooth spectroscopic response at long times. This difference is clearly manifested also in Fig. \ref{spect_strong_exp_theory}(c) which compares the calculated backward mode population at $t = 200 \ \mu $sec as a function of the frequency difference between the pump beams for the fixed phase and the averaged phase excitation. The fixed phase response shows a spectrum consisting of several sharp peaks, which are related to the Fourier broadening of the spectrum (at time $t=200 \mu$sec, this should correspond to peaks at $\sim 5$ kHz). We note that these peaks are enhanced and broadened by the nonlinear nature of the superradiance process. The averaged phase spectrum appears centered around $f \simeq 13$ kHz, with a smoother and rather wide response of $\sim 20$ kHz. The strong-pulse spectrum is strongly broadened and down-shifted as compared to the expected energy-conserving resonance frequency of $4 \omega_R /2 \pi \simeq 15$ kHz (which is recovered in the weak-pulse case presented below). 

This non-trivial shift of the resonance can be intuitively explained by considering the nonlinear dynamics of the process, in which the forward atomic mode grows first, followed by growth of the backward atomic mode. As mentioned earlier, the backward mode resonant coupling is related to the atomic recoil energy. However, the forward mode coupling does not depend on this effect. Rather, the forward modes, which are equally generated by the two frequency components of the pump, interfere and oscillate according to the time-dependent phase difference between them (see Fig. \ref{td_calc_exp}). Therefore, the forward mode population experiences less fluctuations for lower frequency differences. Taking this effect into account, together with the strong coupling compared to the backward mode resonance frequency and the nonlinearity of the growth process, we find that the backward mode resonance frequency is effectively pulled to a lower frequency.

Figure \ref{spect_strong_exp_theory}(d) shows the measured spectroscopic response of the backward peaks for a fixed phase of $\pi$ between the two pump frequencies. The measured spectrum is composed of a broad feature also seen in the calculated spectrum, and a downward shift of the resonance from  $4 \omega_R / 2 \pi \simeq 15$ kHz. 
The experimental results fit the main features of the calculated spectrum. However, we could not resolve the peak structure found theoretically for the fixed phase excitation. We have repeated our measurements for an averaged phase pump, and observed a similar qualitative spectrum. We have also performed measurements and calculations of the forward modes spectrum, and found both were nearly flat and featureless, as expected. 

\begin{figure}[tbh]
\centering
\includegraphics[width=0.78\linewidth,height=0.4\linewidth]{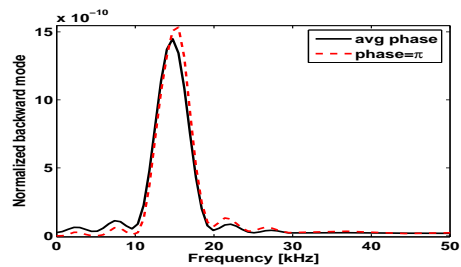}
\caption{Calculation of the backward mode population at $t=200 \ \mu$sec, as a function of the pump beam frequency difference.
These results are for weak coupling, averaged over 10 randomly chosen phases between the frequency components (black-solid) and for a fixed phase of $\pi$ (red-dashed).
A distinct, narrow resonance is visible at $ f = 4 \omega_R /2 \pi \simeq 15$ kHz.}
\label{fcalc_lowg}
\end{figure}
Thus far we presented the strong coupling regime, since for the weak coupling case poor signal-to-noise prevented measurement of the superradiance mode populations. To better understand the results  of strong-pulse superradiance, we study theoretically the effect of weak coupling by reducing the coupling coefficient to $G \simeq 0.06$ (two orders of magnitude smaller than the coupling used experimentally). In Fig. \ref{fcalc_lowg} we calculate the frequency response in the weak coupling regime with a fixed phase and an averaged phase between the two pump frequencies. The spectrum in this limit exhibits a rather narrow resonance, clearly centered around $f = 4 \omega_R / 2 \pi$. This is the expected position of the resonance, taking into account that the backward modes are detuned from the forward modes by this energy difference. In this regime both the fixed phase and averaged phase spectra exhibit a single-peak structure, signifying that for weak coupling coherent oscillations are weak, and the resonant response dominates. Still, small peaks at $\sim 5$ kHz intervals are noticeable, which, as in the strong pulse case, result from the finite time at which the spectrum was taken ($t=200 \mu$sec).

In conclusion, we have studied, both theoretically and experimentally, superradiant excitation of a BEC in the strong coupling regime. We used a pair of collinear pump beams with a controllable frequency difference between them to perform for the first time spectroscopic measurements of the superradiance process, through the population of the backward scattering atomic modes. These backward modes are off-resonant from the single-frequency superradiance by $4 \omega_R / 2 \pi$. We observe a shift of the backward scattering two-frequency resonance to lower frequency in the averaged-phase spectrum, due to the exponential nature of the superradiance process, which is captured by the calculation. In addition, we perform coherent time dynamics, revealing oscillations of the mode populations and a peaked structure of the fixed-phase spectrum. This is in qualitative agreement with the theoretical model, which is semi-classical, and indicates the coherence of the measured dynamics. These novel results open up further possibilities for studying quantum fluctuations and buildup of superradiance in ultracold atomic gases.

After the completion of this work it has come to our knowledge that related two-frequency superradiance experiments were carried out in the weak coupling regime \cite{straten}. This work was funded in part by DIP and the Minerva Foundation.

\bibliography{bibBEC3}


\end{document}